\documentstyle[12pt,epsfig]{article}

\newlength{\dinwidth}
\newlength{\dinmargin}
\begin{document}
\def\bold#1{\setbox0=\hbox{$#1$}%
     \kern-.025em\copy0\kern-\wd0
     \kern.05em\copy0\kern-\wd0
     \kern-.025em\raise.0433em\box0 }
\def\slash#1{\setbox0=\hbox{$#1$}#1\hskip-\wd0\dimen0=5pt\advance
       \dimen0 by-\ht0\advance\dimen0 by\dp0\lower0.5\dimen0\hbox
         to\wd0{\hss\sl/\/\hss}}
\newcommand{\be}{\begin{equation}}
\newcommand{\ee}{\end{equation}}
\newcommand{\bea}{\begin{eqnarray}}
\newcommand{\eea}{\end{eqnarray}}
\newcommand{\nn}{\nonumber}
\newcommand{\dd}{\displaystyle}
\newcommand{\bra}[1]{\left\langle #1 \right|}
\newcommand{\ket}[1]{\left| #1 \right\rangle}
\newcommand{\spur}[1]{\not\! #1 \,}
\thispagestyle{empty}
\vspace*{1cm}
\rightline{BARI-TH/01-418}
\rightline{July 2001}
\vspace*{2cm}
\begin{center}
  \begin{LARGE}
  \begin{bf} 
$D_s$ decays to $\eta$ and $\eta^\prime$ final states:\\
a phenomenological analysis
  \end{bf}
  \end{LARGE}
\end{center}  
\vspace*{8mm}
\begin{center}
\begin{large}
P. Colangelo and F. De Fazio
\end{large}
\end{center}  

\begin{center}
\begin{it}
 Istituto Nazionale di Fisica Nucleare, Sezione di Bari, Italy
\end{it}
\end{center}
\begin{quotation}
\vspace*{1.5cm}
\begin{center}
  \begin{bf}
  Abstract\\
\end{bf}  
\end{center}  
\noindent
We consider the semileptonic and nonleptonic $D_s$ decay modes to final states
with $\eta$ and $\eta^\prime$. We use QCD sum rules to determine the
$D_s \to \eta$ form factor $f_+^\eta$, and a
generalized factorization ansatz to compute nonleptonic decays.
We propose a parameterization of possible OZI suppressed contributions 
producing the $\eta^\prime$  in the final state, compatible with
current data; such a scheme can be further constrained
improving the precision of the measurement
of the $D_s$ decay rates, as expected by the ongoing experiments.
\vspace*{0.5cm}
\end{quotation}

\newpage
\baselineskip=18pt
\vspace{2cm}
\noindent
\section{Introduction}
The exclusive $D_s$ decays to final states containing $\eta$ and $\eta^\prime$
represent nearly $30\%$ of the total decay rate of the  $D_s$ meson.
Therefore, $D_s$ could be a suitable system where to gather information on
important aspects of the $\eta-\eta^\prime$ phenomenology,
namely the long-standing issue of the $\eta-\eta^\prime$ mixing. 
Moreover, $D_s$ can be used to further investigate 
some unsettled  aspects of  nonleptonic heavy meson decays, such as
the  anomalously large $\eta^\prime$ production observed in several 
heavy meson decay 
channels. Examples are $B^- \to K^- \eta^\prime$ and 
$D^0 \to \overline{K}^0 \eta^\prime$, the measured decay rates of which 
are substantially larger than what can be expected by naive theoretical 
calculations. 

The current experimental situation concerning the $D_s$ transitions to $\eta$
and $\eta^\prime$ is summarized in 
Table \ref{tab:tab1} \cite{pdg}, mainly using the results obtained 
by the CLEO Collaboration in the past few years \cite{cleo1}. 
\begin{table}[h]
\caption{Experimental rates and branching fractions 
of semileptonic and nonleptonic $D_s$ 
decays to final states containing $\eta$ and $\eta^\prime$.} 
\label{tab:tab1} \begin{center}
\begin{tabular}{|| l c c ||} \hline \hline
Decay mode & $\Gamma$ $\,(10^{-15}\,GeV)$ &  ${\cal B}(10^{-2})$ \\
\hline
$D_s^+ \to \eta \ell^+ \nu$        &$34.5\pm9.3$ &$2.6 \pm 0.7$ \\
$D_s^+ \to \eta^\prime \ell^+ \nu$ &$11.8\pm4.5$ &$0.89 \pm 0.34$\\
\hline 
$D_s^+ \to \eta \pi^+$        & $22.6\pm6.7$   & $1.7  \pm 0.5 $ \\
$D_s^+ \to \eta \rho^+$       & $143.3\pm41.2$ & $10.8 \pm 3.1 $ \\
$D_s^+ \to \eta^\prime \pi^+$ & $51.8\pm13.3$  & $3.9  \pm 1.0 $ \\
$D_s^+ \to \eta^\prime \rho^+$& $134.0\pm37.3$ & $10.1 \pm 2.8 $ \\
\hline
\hline
\end{tabular}
\end{center}
\end{table}
%
The experimental results are expected to be  improved
in the near future, since the analysis of the $D_s$ system is an important
item of the experimental 
program of the current hadron facilities, as
well as of the $e^+ e^-$ machines running at the $\Upsilon(4S)$ peak. 

The results in Table \ref{tab:tab1} have inspired several considerations.
First, it has been proposed
that information on the $\eta-\eta^\prime$ mixing  could
be obtained just considering the semileptonic decay modes. As a matter
of fact,
writing the hadronic matrix element governing the transition
$D_s^+ \to \eta \ell^+ \nu$ in terms of form factors:
\be                                                               
< \eta(p^\prime) | {\bar s} \gamma_\mu c | D_s(p) >=   
f_+^\eta(q^2) (p+p^\prime)_\mu +                     
f_-^\eta(q^2) q_\mu  \label{matrix}    
\ee                                                               
($q=p-p^\prime$) and  a similar expression for 
$D_s^+ \to \eta^\prime \ell^+ \nu$, the ratio 
$\displaystyle {{\cal B}(D_s^+ \to \eta^\prime \ell^+ \nu) \over
{\cal B}(D_s^+ \to \eta \ell^+ \nu)}$ could be used to access the 
$\eta-\eta^\prime$ mixing angle through the ratios of the form factors
$\displaystyle{f_\pm^{\eta^\prime}(q^2)/ f_\pm^\eta(q^2)}$
which are  related to the $\eta-\eta^\prime$ mixing scheme 
\cite{verma,anisovich}.
In particular, 
information could be gathered on the
mixing scheme in the flavour basis
\cite{feld,feldrev}, which consists in writing
the $\eta$ and $\eta^\prime$ states as  combinations of 
$|\eta_q>= \displaystyle{1 \over \sqrt{2}}|{\overline u}
u+ {\overline d} d>$ and $|\eta_s>=|{\overline s} s>$:
\bea
|\eta> &=& \cos \phi_q ~ |\eta_q> -\sin \phi_s ~|\eta_s> \nonumber \\
|\eta^\prime> &=& \sin \phi_q ~ |\eta_q> +\cos \phi_s~ |\eta_s> \;\;\; .
\label{statemix}
\eea
It has been shown \cite{feld} that in this scheme a single
angle is essentially required, since 
$|\phi_s-\phi_q|/ (\phi_s+\phi_q)\ll 1$, a result 
confirmed by a QCD sum rule calculation \cite{penn}.
Therefore, one can safely
assume $\phi_s \simeq \phi_q \simeq \phi$; the most
recent estimates of $\phi$ give values close to  $40^0$
\cite{feldrev,escribano}.
In the flavour scheme, the semileptonic form factors relative to 
$D_s^+ \to \eta \ell^+ \nu$ and $D_s^+ \to \eta^\prime \ell^+ \nu$  satisfy
the relation
\be
{|f_\pm^{\eta^\prime}(q^2)| \over |f_\pm^\eta(q^2)|}=\cot \phi \;\;\; ,
\label{phiff}
\ee
so that the possibility of  a direct comparison with the results for
$\phi$ obtained 
from the analyses of other channels involving  $\eta-\eta^\prime$
particles could be envisaged.
The situation is particularly simple in the case of 
semileptonic $D_s^+$ decays to positrons or antimuons, where essentially 
only the form factors $f_+^{\eta(\eta^\prime)}(q^2)$ are involved.
However, in order to pursue this program, one has to neglect possible
contributions to the semileptonic decay amplitude
from diagrams where $\eta$ and $\eta^\prime$ are produced  
through  gluon emission; we shall consider this problem below. 

As for nonleptonic decays, naive factorization,
using the semileptonic $D_s \to \eta$ and $D_s \to \eta^\prime$
form factors and the Wilson coefficients relevant for
the transitions in Table \ref{tab:tab1}, does not allow
to  predict all the branching fractions
of $D_s^+ \to \eta^{(\prime)} \pi^+$ and $D_s^+ \to \eta^{(\prime)} \rho^+$
\cite{oldattempts}.
The same conclusion is obtained by analyzing 
the various decay channels in terms of transition amplitudes related
by $SU(3)_F$ symmetry to analogous amplitudes  for $D$ decays
\cite{rosner},
or accounting for some  effects of the inelastic final state rescattering
\cite{pham}. In particular, the prediction
for the rate of the decay mode $D_s^+ \to \eta^\prime \rho^+$ is lower 
than the experimental measurement by more than a factor of two.
This is disappointing: in
Cabibbo favoured hadronic $D_s$ decays the final state 
contains a single isospin mode, thus ruling out possible interference effects 
due to the elastic final state interactions;
 moreover, the conservation of G-parity does not allow to include inelastic
effects of intermediate states consisting of ordinary mesons in the
$D_s^+ \to \eta^\prime \rho^+$ mode
\cite{lipkin}. Therefore, a different mechanism must be invoked to explain
the enhanced $\eta^\prime$ production.  
It has been suggested that the enhancement could be due to  OZI
suppressed diagrams with
the $\eta'$ produced by gluons and the $c \bar s$ pair annihilating
to a charged $W$ \cite{verma}. 
This mechanism would not affect substantially the $\eta$ production, since
the coupling of the gluons to $\eta$ is estimated to be smaller than
the coupling to $\eta^\prime$  \cite{novikov}. However, a mechanism
of this type, violating the OZI rule, could also affect  
the semileptonic $D_s^+ \to \eta^\prime \ell^+ \nu$ transition, 
spoiling  the  possibility of using the 
relation (\ref{phiff}) to gather information on the angle $\phi$ from the 
semileptonic decay rates.
Moreover, these effects could be also present in other systems, namely
in $D$ decays,  although in such cases the annihilation amplitudes are 
Cabibbo suppressed.

The effects of the  gluon production of the $\eta$ and 
$\eta^\prime$,
although plausible,  are not included in ordinary
analyses since they are  difficult to take into account in a quantitative way.
Nevertheless, their investigation is  of particular relevance, 
and we shall try to perform it in a phenomenological way. 

In this paper, we  compute the form factor $f_+^\eta(q^2)$ relative to 
$D_s^+ \to \eta \ell^+ \nu$,
showing that the result is in agreement with the experimental measurement
in Table \ref{tab:tab1}. On the other hand, 
assuming the standard value of the $\eta - \eta^\prime$ mixing angle
together with the naive factorization, other 
results in Table \ref{tab:tab1} are not reproduced.
Therefore,  we adopt a generalized 
factorization ansatz, fitting the relevant parameters  from the
experiment; 
moreover, we assume that the effect of the process producing the $\eta^\prime$
through the annihilation of the $c \bar s$ pair numerically
modifies the $D_s \to \eta^\prime$ form factors. This enables us to
investigate whether the experimental results can be reproduced by
this assumption 
and how the reduction of the experimental uncertainty can be used to 
test various consequences of our ansatz.

\section{QCD sum rule calculation of  $f_+^\eta(q^2)$}

\begin{figure}[ht]
\begin{center}
\mbox{\epsfig{file=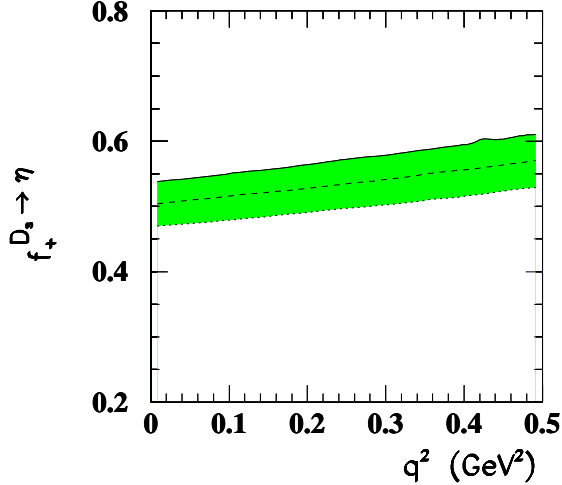, width=7.5cm}}
\end{center}
\caption{Form factor $f_+^{D_s \to \eta}(q^2)$ as obtained using QCD sum
rules.
The shaded region represents the theoretical uncertainty related to the
variation
of the input parameters.}
\label{figure1}
\end{figure}

Let us first compute the form factor $f_+^\eta(q^2)$ using a nonperturbative
method, such as the QCD sum rule technique \cite{shifman}.      
We  adopt the usual strategy of considering a three-point function:   
\bea                                                                          
\Pi_{\mu \nu}(p^2,p^{\prime 2},q^2)&=&i^2 \int d^4 x~ d^4 y~ e^{-i p \cdot
y}\,   
e^{i p^\prime \cdot x} \bra{0} T[J^\eta_5(x) J_\mu(0) J_5^{D_s}(y)]\ket{0}
\nonumber \\
&=& \Pi_+ P_\mu+ \Pi_- q_\mu \;\;\; ,
\label{cor} 
\eea
with $J^\eta_5={\bar s }i \gamma_5 s$ the pseudoscalar quark density probing
the strangeness content of the $\eta$, $J_\mu={\bar s }\gamma_\mu c$ the
weak current inducing the $c \to s$ transition, and 
$J_5^{D_s}={\bar c}i \gamma_5 s$  a quark current having the $D_s$
quantum numbers. The momenta $P$ and $q$ are defined as $P=p+p^\prime$
and $q=p-p^\prime$, respectively. For the invariant function
$\Pi_+(p^2,p^{\prime2},q^2)$ 
a double dispersion relation in  the variables
$p^2$, $p^{\prime 2}$ can be written down:
\be
 \Pi_+(p^2,p^{\prime 2},q^2)={1 \over \pi^2} \int ds_1 \int ds_2\,
{\rho(s_1,s_2,q^2) \over (s_1 -p^2)(s_2-p^{\prime 2})} 
\ee
where possible subtraction terms have been omitted.
The spectral function $\rho(s_1,s_2,q^2)$ contains,
for low values of $s_1$ and
$s_2$, a double $\delta-$function corresponding to the transition $D_s
\to \eta$. Isolating such a  contribution, and neglecting  possible
subtraction terms which we discuss later on,
 we can write:
\be
\Pi_+(p^2,p^{\prime 2},q^2)={{\cal A} f_+^\eta(q^2) \over
(M_{D_s}^2-p^2)(M_\eta^2-p^{\prime 2})} {f_{D_s} M_{D_s}^2 \over m_s+m_c}+
{1 \over \pi^2} \int_{s_1^0}^\infty
ds_1 \int_{s_2^0}^\infty d s_2\, {\rho^{had}(s_1,s_2,q^2) \over (s_1
-p^2)(s_2-p^{\prime 2})} \; . \label{hadr}
\ee 
In (\ref{hadr}) we have assumed that the contribution of higher
resonances and continuum of states starts from the effective thresholds 
$s_1^0$ and
$s_2^0$. The hadronic parameter $\cal A$ represents the matrix element:
\be
< 0| J_5^\eta |\eta (p^\prime) > = {\cal A}  \label{defa}
\ee
while the projection of the $J^{D_s}_5$ current on the $D_s$ state is
given by the matrix element 
 \be
< 0|J_5^{D_s} |D_s(p) > = {f_{D_s} M_{D_s}^2 \over m_s+m_c} \;\;\; .
\label{matrelem}
\ee
The correlator (\ref{cor}) can  be computed in QCD for large Euclidean
values of $p^2$ and $p^{\prime 2}$ by an
Operator Product Expansion, 
expanding the T-product in (\ref{cor}) as a sum of a perturbative 
contribution
and non perturbative terms,  proportional to vacuum expectation
values of quark and gluon gauge invariant operators of increasing
dimension, the vacuum condensates. In practice, only the first few
condensates numerically contribute, the most important ones being the
dimension 3 $<{\bar s} s>$ and dimension 5 
$<{\bar s} g \sigma G s>$ condensates. 
The QCD expression for $\Pi_+$ reads:
\bea
\Pi_+(p^2,p^{\prime 2},q^2)&=&{1 \over \pi^2}\int_{(m_c+m_s)^2}^\infty d s_1
\int_{4
m_s^2}^\infty d s_2
{\rho^{pert}(s_1,s_2,q^2) \over (s_1 -p^2)(s_2-p^{\prime 2})}\nonumber \\
&+&\Pi_+^{(D=3)} <{\overline s} s> 
+\Pi_+^{(D=5)} < {\overline s} g \sigma G s> +...\;\;.
\label{opexp}
\eea
\noindent 
Invoking quark-hadron duality, i.e. assuming that the
hadronic and the perturbative QCD spectral densities give the same result when
integrated above the thresholds $s_1^0$ and $s_2^0$,  we get the  sum
rule:
\bea
{{\cal A} f_{D_s} M^2_{D_s} \over m_s+m_c} {f_+^\eta (q^2) \over
(M_\eta^2-p^{\prime 2}) (M_{D_s}^2-p^2)}
&=&
{1 \over 4 \pi^2} \int_D d s_1 d
s_2 {\rho_+^{pert}(s_1,s_2,q^2) \over (s_1- p^2)(s_2 -p^{\prime 2})}
\label{fsr} \\
&+&
\Pi_+^{(D=3)}<{\overline s}s >+\Pi_+^{(D=5)}< {\overline s} g \sigma G s>
+... 
\nonumber \eea
with
\bea
\rho_+^{pert}(s_1,s_2,q^2)&=&{3 \over 4 \sqrt{(s_1+s_2-q^2)^2-4 s_1 s_2}}
\Big[s_1-m_c^2+m_s^2+s_2+2m_s(m_c-m_s)
\label{rhopert} \\
&+& 2s_2{(s_1+s_2-q^2)(2s_1-m_c^2+m_s^2) \over
(s_1+s_2-q^2)^2-4s_1 s_2}
-{(s_1+s_2-q^2)^2(s_1-m_c^2+m_s^2+s_2) \over (s_1+s_2-q^2)^2-4s_1
s_2} \Big]
\nonumber 
\eea
and
\bea
\Pi_+^{(D=3)}&=& - {1 \over 2}
\Big\{ {m_c \over r r^\prime}
\label{d=3} \\
&+&(m_c+m_s) \Big[{m_s^2 \over 2 r r^{\prime
2}}+{m_c m_s \over r^2 r^\prime}
-{m_s^2 (m_c^2+m_s^2-q^2) \over 2 r^2
r^{\prime 2}}
+m_s^2 \Big({m_s^2 \over r r^{\prime 3}}+{m_c^2 \over r^3
r^\prime} \Big) \Big] \Big\}
\nonumber 
\eea
\bea
\Pi_+^{(D=5)}&=&{1\over 24}
\Big[ {6 m_s^2(m_c+ m_s) \over r r^{\prime 3 }} +{6 m_c^2(m_c+ m_s) \over
r^3 r^{\prime }} \\
&+&{2 (m_c+ m_s) (2 m_c^2+2 m_s^2-2 q^2-m_c m_s) \over r^2
r^{\prime 2}}+{12 m_c \over r^2 r^\prime}+{12 m_s \over r r^{\prime 2}} \Big]
\,.\nonumber
\eea
The variables $r$ and $r^\prime$ are defined as $r=p^2-m_c^2$ and
$r^\prime=p^\prime-m_s^2$.
The domain $D$ is bounded by the curves 
\bea
s_2^\pm &=& {[2 m_s^2 (s_1+q^2)+\Delta (m_c^2-m_s^2-q^2)] \over 2
m_c^2} \label{domain}\\
& \pm & {1 \over 2 m_c^2} \left[ \left( 2 m_s^2(s_1+q^2)+\Delta
(m_c^2-m_s^2-q^2) \right)^2 -4 m_c^2 m_s^2 (s_1-q^2)^2 \right]^{1/2} 
\nonumber
\eea
where $\Delta=s_1-m_c^2+m_s^2$, and by the lines $s_2=s_2^0$ and
$s_1=s_1^0$.
Eq. (\ref{fsr}) can be improved by applying
to both its sides a  Borel transform, defined as follows:
\be
{\cal B} [{\cal F}(Q^2)]=lim_{Q^2 \to \infty, \; n \to \infty, \; {Q^2
\over
n}=M^2}\;
{1 \over (n-1)!} (-Q^2)^n \left({d \over dQ^2} \right)^n {\cal F}(Q^2) \;
,
\label{tborel}
\ee
where ${\cal F}$ is a generic function of $Q^2$. The application of
such a procedure to the sum rule amounts to exploiting the  relation:
\be
{\cal B} \left[ { 1 \over (s+Q^2)^n } \right]={\exp(-s/M^2) \over
(M^2)^n\ (n-1)!} \; , \label{bor}
\ee
with $M^2$ a Borel parameter. The operation,
applied independently to the variables $-p^2$ and $-p^{\prime 2}$, 
improves the
convergence of the series in the OPE in the r.h.s. of
Eq. (\ref{opexp}) by factorials in $n$, and, for
suitably chosen values of the Borel parameters, enhances the contribution
of the low-lying states in the hadronic representation
of the correlator $\Pi_+$. Moreover,
since the Borel transform of a polynomial vanishes, the procedure allows
to get rid of  subtraction terms in the dispersion relations, which are
polynomials in $p^2$ or $p^{\prime 2}$. Therefore, a final sum rule can be 
worked out, keeping only the contribution of the lowest dimensional 
condensates:
\bea
{{\cal A} f_{D_s} M^2_{D_s} \over m_s+m_c} f_+^\eta (q^2) 
 e^{-{M_{D_s}^2/M_1^2}} e^{-{M_\eta^2/M_2^2}} 
&=&
{1 \over 4 \pi^2} \int_D d s_1 d
s_2 \rho_+^{pert}(s_1,s_2,q^2) 
e^{-{s_1 \over M_1^2}} e^{-{s_2 \over M_2^2}} 
\nonumber \\
-e^{-{m_c^2 / M_1^2}} e^{-{m_s^2 / M_2^2}} 
{<{\bar s}s> \over 2}
\Big\{m_c &+& (m_c+m_s) \Big[-{m_s^2 \over 2 M_2^2}-{m_c m_s \over 2
M_1^2}
\nonumber \\
&-&{m_s^2(m_c^2+m_s^2-q^2) \over 2 M_1^2 M_2^2}+{m_s^2 \over 2 }
 \Big({m_s^2 \over M_2^4}+{m_c^2 \over M_1^4} \Big) \Big] \Big\}
\nonumber \\
+e^{-{m_c^2 / M_1^2}} e^{-{m_s^2 / M_2^2}} 
{<{\bar s} g \sigma G s> \over 8}  \Big[&&{m_s^2 (m_c+m_s) \over
M_2^4}+{m_c^2 (m_c+m_s) \over M_1^4}
\nonumber \\
&+&{2 (m_c+m_s) (2 m_c^2+2 m_s^2-2 q^2-m_c m_s) \over 3 M_1^2 M_2^2}
\nonumber \\
&-&{4 m_c \over M_1^2}-{4 m_s \over  M_2^2} \Big] \,.
\label{borelsr}
\eea
In the numerical analysis of (\ref{borelsr}) we used standard values of the
condensates:
$<{\bar s}s>=0.8<{\bar q}q>$ with $<{\bar q}q>=(-0.23\; GeV)^3$, and 
$<{\bar s}g \sigma G s>= m_0^2 <{\bar s}s>$ with  $m_0^2=0.8 \; GeV^2$.
The charm and strange quark masses were fixed  to the values 
$m_c=1.4 \, GeV$ \cite{mcharm} and
$m_s=140\,MeV$  \cite{ms,mcharm}. As for the $D_s$ decay constant, we used
$f_{D_s}=225\,MeV$ \cite{mcharm}, while for the parameter $\cal A$ we adopted
the
two-point QCD sum rule result ${\cal A}=0.115\, GeV^2$ \cite{penn}.
The obtained sum rule shows stability to the variation of the
Borel parameter in the region $2.5 \; GeV^2 \le M_1^2 \le 3.5 \; GeV^2$ 
and $1.6 \; GeV^2 \le M_2^2 \le 2.4 \; GeV^2$, with the thresholds 
$s_1^0$ and $s_2^0$ in the ranges $s_1^0=5.9-6.1$ $GeV^2$
and $s_2^0=0.9-1.1$ $GeV^2$, respectively.
The form factor $f_+^\eta(q^2)$, obtained in the range of momentum transfer
$0 \le q^2 \le 0.5$ $GeV^2$, 
is depicted in Fig. \ref{figure1}; 
it can be fitted by a linear expression 
\be
f_+^\eta(q^2)=A \,q^2 + B \label{fpiu} \, ,
\ee
with $A=0.14$ $GeV^{-2}$ and $B=0.50 \pm 0.04$.
This expression is consistent, in the considered range of momentum transfer, 
with a polar form  $ \displaystyle{f_+^\eta(q^2)={f_+^\eta(0) \over 1-{q^2
\over M_P^2}}}$, with the mass of the pole $M_P \simeq 1.9$  $GeV$.

In the following, we shall consider the  form factor $f_+^\eta(q^2)$
as a theoretical input in a 
phenomenological analysis of $D_s$ transitions.
\section{$D_s$ transitions to $\eta$ and $\eta^\prime$}

The form factor $f_+^\eta(q^2)$ computed above allows us
to calculate the semileptonic $D_s^+ \to \eta \ell^+ \nu$ decay rate.  It
can also be used to analyze the
nonleptonic  modes $D_s \to \eta \pi^+$ and $D_s \to \eta \rho^+$
if the factorization approximation 
is adopted. This amounts to consider the effective Hamiltonian 
\be
H_{eff}
={G_F \over \sqrt{2}}V_{cs}^*V_{ud} \left(C_1(\mu)+ {C_2(\mu)\over 
N_c}\right)
({\bar s}c)_{V-A}({\bar u}d)_{V-A}+ \, h.c. \label{heff} \,,
\ee
with $({\bar q}_1 q_2)_{V-A}={\bar q}_1 \gamma_\mu (1-\gamma_5) q_2$ 
and  $C_1$ and $C_2$  Wilson coefficients, and factorize the $V-A$
currents appearing in it.
As for the modes with $\eta^\prime$,
we further need an input on the $\eta-\eta^\prime$ mixing, and we choose 
the angle $\phi$ in the flavour basis mixing scheme, with
the value $\phi=39^0$ coming from the
measurements of  $\phi \to \eta^{(\prime)} \gamma$
\cite{novosibirsk}.
\begin{table}[h]
\caption{Computed semileptonic and nonleptonic $D_s$ rates and
branching fractions. Nonleptonic rates are obtained using naive 
factorization. The
$\eta-\eta^\prime$ mixing is described in the flavour basis, with mixing
angle
$\phi=39^0$. } \label{brs} \begin{center}
\begin{tabular}{|| l c c ||} \hline \hline
Decay mode & $\Gamma$ $\,(10^{-15}$ GeV) & ${\cal B} \,(10^{-2})$ \\
\hline
$D_s^+ \to \eta \ell^+ \nu$        & $30.3\pm 4.8$&$2.3\pm 0.4$\\
$D_s^+ \to \eta^\prime \ell^+ \nu$ & $12.7\pm 2.0$&$1.0\pm 0.2$\\
\hline \hline
$D_s^+ \to \eta \pi^+$        &$38.5\pm 6.2 $&$2.9\pm 0.5$\\
$D_s^+ \to \eta \rho^+$       &$74.5\pm 11.9$&$5.6\pm 0.9$\\
$D_s^+ \to \eta^\prime \pi^+$ &$33.2\pm 5.3 $&$2.5\pm 0.4$\\
$D_s^+ \to \eta^\prime \rho^+$&$30.7\pm 4.9 $&$2.3\pm 0.4$\\
\hline
\hline
\end{tabular}
\end{center}
\end{table}
In Table \ref{brs} we collect the resulting  branching fractions 
obtained in the factorization approximation, using
$f_\pi=0.132$ $GeV$, $f_\rho=0.220$ $GeV$, $\tau_{D_s}=0.496 \; ps$; 
the number of colours $N_c$ is fixed to $N_c=3$, and the values
$C_1(m_c)=1.263$ and $C_2(m_c)=-0.513$ are chosen,  corresponding to
the results for  the Wilson coefficients
obtained at the leading order in renormalization group 
improved perturbation theory at $\mu=m_c\simeq 1.4~GeV$, in corrispondence
to $\alpha_s(M_Z)=0.118$.
Using the form factor $f_+^\eta$ in (\ref{fpiu}) we obtain the branching 
fraction ${\cal B}(D_s^+ \to \eta \ell^+ \nu)=(2.3 \pm 0.4)\times 10^{-2}$
in agreement with the experimental outcome reported in Table
\ref{tab:tab1}; also the result 
${\cal B}(D_s^+ \to \eta^\prime \ell^+ \nu)=(1.0 \pm 0.2)\times 10^{-2}$,
obtained using Eq.(\ref{phiff}),
is within the experimental uncertainty quoted in Table \ref{tab:tab1}.
On the other hand, as one can infer by comparing the computed decay rates
reported in Table \ref{brs} with the experimental measurements in
Table \ref{tab:tab1}, the calculations of the nonleptonic modes
do not fit all the experimental measurements, as already anticipated by
previous analyses. 

In order to parameterize the deviation
from the factorization approximation, as well as the possible role of 
the $\eta$ and $\eta^\prime$  gluon production, we adopt  
a generalized factorization ansatz, 
consisting in substituting the combination of the
Wilson coefficients $a_1=C_1 + \displaystyle{C_2 \over N_c}$ with
effective scale-independent parameters $a_1^{eff}$ in the 
factorized amplitudes. The coefficients  $a_1^{eff}$ should be considered as 
non-universal, process-dependent  parameters \cite{neubstech}. However,
since 
in the decay modes $D_s^+ \to \eta \pi^+$, $D_s^+ \to \eta^\prime \pi^+$,
 and analogously  $D_s^+ \to \eta \rho^+$, $D_s^+ \to \eta^\prime
\rho^+$, the underlying process is the same, we
assume only two process-dependent parameters to describe
the deviation from naive factorization: $a_{1,\pi}^{eff}$ describing  
 $D_s^+ \to \eta \pi^+$ and $D_s^+ \to \eta^\prime \pi^+$,  and
$a_{1,\rho}^{eff}$ describing $D_s^+ \to \eta \rho^+$ and 
$D_s^+ \to \eta^\prime\rho^+$. 

As for the possible contribution of OZI suppressed diagrams producing
$\eta$ and $\eta^\prime$, it is essentially related to the matrix
elements 
$\bra{0} G {\tilde G} \ket{\eta^{(\prime)}}$, where $G$ is the gluon field
and $\tilde G$ its dual. 
Several theoretical investigations suggest that 
$\bra{0} G {\tilde G} \ket{\eta} \ll \bra{0} G {\tilde G}
\ket{\eta^{\prime}}$ \cite{novikov,penn};  therefore, we assume that
such annihilation amplitudes mainly affect the $D_s$ transitions
to $\eta^\prime$.
A simple parameteterizion consists in modifying
the values of the parameters $A,B$ in (\ref{fpiu}), thus without affecting the
shape of the form factor $f_+^{\eta^\prime}$. This seems rather reasonable,
 since the range of momentum transfer 
in $D_s \to \eta^\prime$ transitions is rather narrow 
($q^2\simeq 0$ for $D_s \to \eta^\prime \pi$, 
$q^2= M^2_\rho$ for $D_s \to \eta^\prime \rho$ and 
$q^2 < (M_{D_s}-M_{\eta^\prime})^2$ for $D_s \to \eta^\prime \ell \nu$),
and a linear expansion is a suitable representation of the
form factors. Therefore, in the case of $\eta^\prime$, we phenomenologically 
represent the $D_s \to \eta^\prime$ form factor as
\be
f_+^{eff}(q^2)={\bar A} q^2 + {\bar B} \label{fpiueff} \,.
\ee

It is now possible  to use the experimental data in
Table \ref{tab:tab1} to fit all the parameters 
we have introduced, namely $a_{1,\pi}^{eff}$, $a_{1,\rho}^{eff}$,
$\bar A$ and $\bar B$.
From the decays 
$D_s^+ \to \eta \pi^+$ and $D_s^+ \to \eta \rho^+$ we find that the
values of $a_{1,\pi}^{eff}$ and $a_{1,\rho}^{eff}$ are bound in
the ranges:
\be
a_{1,\pi}^{eff} \in [0.65,1.04] \hskip 1.5 cm 
a_{1,\rho}^{eff} \in [1.21,1.86] \label{aieff} \,,
\ee
to be compared with the value of $a_1$ obtained from the Wilson coefficients
$C_1$ and $C_2$:
$\displaystyle{a_1(m_c)=C_1(m_c)+{C_2(m_c) \over
N_c}\simeq 1.1}$.
As for the decay mode $D_s^+ \to \eta^\prime \pi^+$, 
it involves $f_+^{eff}(q^2)$; however,
only the value
$f_+^{eff}(0)={\bar B}$ is needed in the approximation $M_\pi =0$,
allowing us to  constrain ${\bar B}$ in the range
\be
|{\bar B}| \in [0.70,1.45] \label{bbar} \,.
\ee
Moreover, considering the modes $D_s^+ \to \eta^\prime \ell^+ \nu$ and 
$D_s^+ \to  \eta^\prime   \rho^+$, we find that the relations   
\bea 
\Gamma (D_s^+ \to \eta^\prime \rho^+)&=& X_\rho \left[ f_+^{eff}(M_\rho^2)  \right]^2  \left[ a_{1,\rho}^{eff} \right]^2 
\label{sistema1} \\ 
\Gamma(D_s^+ \to \eta^\prime \ell^+ \nu) 
&=& X_{semilep}
\int_0^{(M_{D_s}-M_{\eta^\prime})^2} dq^2 \left[ \lambda(M_{D_s}^2,
M_{\eta^\prime}^2, q^2) \right]^{3/2} \left[ f_+^{eff}(q^2)\right]^2  
\label{sistema2} 
\eea 
 ($X_\rho=\displaystyle{G_F^2 |V_{ud}V_{cs}^*|^2f_\rho^2 \over 32 \pi
M_{D_s}^3}\lambda(M_{D_s}^2, M_\rho^2,M_{\eta^\prime}^2)^{3/2}$ 
and
$X_{semilep}=\displaystyle{G_F^2 V_{cs}^2 \over 192 \pi^3 M_{D_s}^3}$)
constrain the parameters
${\bar A}$ and ${\bar B}$ in selected regions of the
$({\bar B},{\bar A})$ plane. These regions are delimited by two
straight lines, from the datum on the nonleptonic $D_s^+ \to \eta^\prime
\rho^+$
decay rate, and  by two ellypses corresponding to the 
measurement of the semileptonic $D_s^+ \to \eta^\prime \ell^+
\nu$ decay rate.

\begin{figure}[ht]
\begin{center}
\epsfig{file=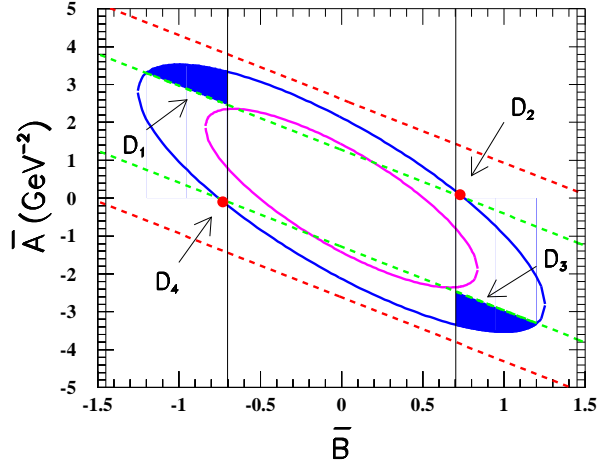, width=8cm}
\end{center}
\caption{Bounds on the parameters $({\bar B},{\bar A})$ in (\ref{fpiueff}). 
The ellypses represent the curves obtained
from Eq. (\ref{sistema2}); the dashed lines stem from Eq. (\ref{sistema1}); 
the two pairs of continuous
vertical lines represent the bound (\ref{bbar}). The shaded areas and the dots
indicate the regions of the parameter space satisfying all the
constraints.
}
\label{figure2}
\end{figure}

Considering simultaneously the constraints, all the data in 
Table \ref{tab:tab1} can be fitted if,
in the ($\bar B, \bar A$) plane,  overlap regions exist among the area
delimited by the ellypses from Eq.(\ref{sistema2}), 
the regions delimited by the straight lines
from Eq.(\ref{sistema1}) and the regions between the vertical lines from
Eq. (\ref{bbar}).  At the present level
of accuracy of the experimental data in Table \ref{tab:tab1} such
regions indeed exist. They  are
depicted in figure \ref{figure2} and denoted as $D_1$, $D_2$, $D_3$ and $D_4$. 
The regions
$D_1$ and $D_3$  are defined, respectively, by the conditions:
\bea
-1.2 \le &{\bar B}& \le -0.70 \label{region2} \\
2.9 \, [0.44-0.58 {\bar B}] \, GeV^{-2}
\le & {\bar A}& \le
1.38 \left[ -1.62 {\bar B}+1.224 \sqrt{1.57 -{\bar B}^2} \right]
GeV^{-2}\nonumber
\eea
\noindent and
\bea
0.70 \le &{\bar B}& \le 1.20 \label{region1} \\
-1.38 \left[ 1.62 {\bar B}+1.224 \sqrt{1.57 -{\bar B}^2} \right]
GeV^{-2}
\le & {\bar A}& \le
-2.9 \, [0.44+0.58 {\bar B}] \, GeV^{-2}\,.
\nonumber
\eea
On the other hand, the regions $D_2$ and $D_4$ are defined, respectively, 
by the conditions:
\bea
0.70 \le &{\bar B}& \le 0.755 \label{region3} \\
2.9\left[ 0.44 - 0.58 {\bar B}\right]  GeV^{-2}
\le & {\bar A}& \le 
1.38 \, [-1.62 {\bar B} + 1.224 \sqrt{1.57 - \bar B^2}] \, GeV^{-2}
\nonumber 
\eea
and
\bea
-0.755 \le &{\bar B}& \le -0.70 \label{region4} \\
-1.38 \, [1.62 {\bar B} + 1.224 \sqrt{1.57 - \bar B^2}] \, GeV^{-2}
\le & {\bar A}& \le 
-2.9 \, [0.44 + 0.58 {\bar B}] \, GeV^{-2}\,.
\nonumber 
\eea

\begin{figure}[ht]
\begin{center}
\mbox{\epsfig{file=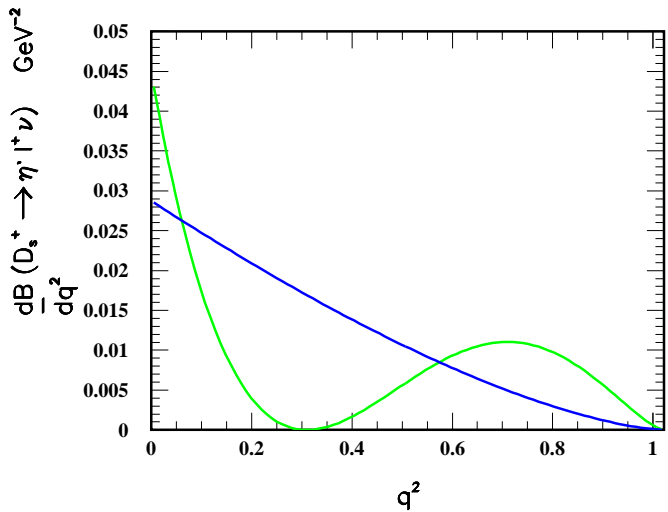, width=9cm}}
\end{center}
\caption{Semileptonic spectra of $D_s \to \eta^\prime \ell \nu$.
The green  curve  corrisponds to the parameters 
$({\bar B},{\bar A})=(0.9,-2.9) \in D_1$, the blue one  to  
$({\bar B},{\bar A})=(0.72, 0.08) \in D_2$.}
\label{spettri}
\end{figure}

Although it is expected and rather plausible, 
the existence of such overlap regions 
was not guaranteed {\it a priori}; it shows that we have chosen a sensible
scheme to parameterize the decays in Table \ref{tab:tab1}. More important,
we expect that an improvement in the accuracy of the experimental data
on the $D_s$ decay rates would sensibly reduce the size of such overlap
regions, and presumably,  exclude some of them. 
Noticeably, already at the present level of accuracy some interesting
observations can be drawn. Let us consider, for example, the parameters in
the regions $D_3$ and $D_2$. In both the cases the experimental
branching fraction
of the semileptonic decay mode $D_s \to \eta^\prime \ell \nu$ is reproduced.
However, a prime difference is that in the region $D_3$ the parameters 
$\bar A$ and $\bar B$  are opposite in sign, while in  the region 
$D_2$ they have the same sign. 
This implies that the relation between the $D_s \to \eta^\prime$ and 
$D_s \to \eta$ form factors in 
(\ref{phiff})  cannot be satisfied by the parameters in the region $D_3$.
The same conclusion holds for the region $D_1$. 
The opposite signs between $\bar A$ and $\bar B$, as it happens in the regions
$D_1$ and $D_3$, have an observable consequence in the spectrum of the 
semileptonic
decay $D_s^+ \to \eta^\prime \ell^+ \nu$:
in this case, a zero in the
$\displaystyle{d \Gamma (D_s \to \eta^\prime \ell \nu) / dq^2}$ 
distribution should be observed, as depicted in Figure
\ref{spettri}. On the other hand,
in the case of parameters in the region $D_2$ (and $D_4$)
a smooth decrease in the spectrum should be observed  as  
in $D \to K \ell \nu$.
%
At the present level of accuracy of the experimental measurements, 
no choice can be done between the two shapes of the semileptonic distribution.

We can reasonably expect that improved data would restrict the allowed 
regions in the $({\bar B},{\bar A})$ plane.
It could happen  
that they do not intersect any more, or that
intersection regions could be found
with restricted extension, allowing a better determination of the
effective parameters introduced in our analysis. 
The calculation of
such parameters remains a challenging task, 
and we do not attempt it in the present paper. However, it is worth outlining 
the theoretical framework in which the calculation could be 
carried out.

Concerning the effective coefficients 
$a_{1,\pi}^{eff}$ and $a_{1,\rho}^{eff}$, which take into account the deviation
from the naive factorization in the corresponding decay modes, their  
theoretical
calculation would consist in  a precise determination of nonfactorizable 
contributions. A step in this direction has been recently performed
in the case of some two-body nonleptonic B decays,  
where the meson picking up the B spectator quark is light,
exploiting the large value of the beauty quark mass \cite{bbns}. In this
case, it has been observed that nonfactorizable contributions are of order 
$\alpha_s$ or $1/m_b$, and a QCD factorization formula has been written for
the nonleptonic matrix elements in terms of meson light-cone distribution 
amplitudes.  A possible extension of such a procedure to charm  requires the
development of a reliable method for computing at least the first (process
dependent) $1/m_Q$ correction. A different approach would 
consist in considering the corrections to the large $N_c$ limit, where
factorization becomes exact \cite{buras}: also in this case, however, 
next-to-leading $1/N_c$ terms are generally sizeable, and one needs their 
actual calculation. Therefore, 
it seems worth attempting to gain information on the effective coefficients
from phenomenological analyses, as done, for example, in
\cite{neubstech}.

As for the OZI suppressed diagrams producing the $\eta^\prime$ through 
its coupling to the gluons, together with a weak annihilation of
$D_s$, a perturbative calculation could be carried out in QCD, in analogy
with the calculation of the $\eta^\prime$ production in 
quarkonium decays \cite{pqcd,quarkonium}. 
The difference, in the present case, is that one
has  to account also for the gluon emission from a light (strange) quark,
and one cannot exploit the fact that all the quarks involved are heavy, 
which justifies the application of perturbative QCD methods.
The calculation,
for small values of $q^2$, produces an amplitude for 
$D_s \to \eta^\prime \ell \nu$ of the same form as provided by a linear 
$q^2$ representation of the $D_s - \eta^\prime$ form factor.
An important ingredient in this perturbative calculation
is the actual value of the two-gluon-$\eta^\prime$
matrix element $<g(k_1) g(k_2)| \eta^\prime(p)>$ describing the vertex
$\eta^\prime g g$ for off-shell gluons. Such a matrix element is 
parameterized by a form factor $F(k^2_1,k^2_2)$ whose value at 
$k^2_1=k^2_2=0$ is fixed by the QCD anomaly;
as for the momentum dependence, various parameterizations have been proposed 
in the literature, thus providing different values for the 
effective parameters $\bar A$ and $\bar B$ introduced in our analysis, which
in turn could correspond to various solutions for the spectrum
shown in fig.\ref{spettri}. One might notice some analogies with
the analyses which explain the observed enhancement  
of the $\eta^\prime$ production in $B$ decays
through the mechanism of gluon fusion \cite{ali}.

All such considerations taken into account, 
we believe that our proposed scheme, where
additional contributions are reabsorbed in the parametrization
of the $D_s \to \eta^\prime$ form factor and in
$a_{1,\pi}^{eff}$, $a_{1,\rho}^{eff}$
is useful from the phenomenological point of view, 
as a starting point for the investigation of the underlying dynamics, and 
could be extended to other cases.

Before concluding, we want to mention  a check of consistency.
If we consider the decay mode $D_s^+ \to {\bar K}^0 K^+$, which can be
related to  $D_s^+ \to \eta \pi^+$ through $SU(3)_F$ symmetry, and
describe the $D_s \to {\bar K}^0$ form factor by $f_+^\eta(q^2)$, 
together with $f_K=0.160\,GeV$, we can estimate the effective parameter 
$a_{1,K}^{eff}$. The experimental measurement 
${\cal B}(D_s^+ \to {\bar K}^0 K^+)=(3.6 \pm 1.1)\times 10^{-2}$ 
produces $a_{1,K}^{eff} \in [0.72,1.14]$, i.e. the effective parameter 
$a_{1,K}^{eff}$ displays a significant overlap with the range
determined for $a_{1,\pi}^{eff}$. In different words, from our analysis and
assuming $SU(3)_F$, we would be able to
predict rather accurately  the experimental datum for ${\cal B}(D_s \to
{\bar K}^0 K^+)$.

\section{Conclusions}

We have presented a phenomenological analysis of
the $D_s$ decays to final states containing $\eta$ and $\eta^\prime$. 
Since  the theoretical investigations 
based on $SU(3)_F$ symmetry, FSI effects and standard $\eta-\eta^\prime$
mixing failed in simultaneously 
reproducing the observed branching ratios for all these decays, 
we have considered a possible role of annihilation diagrams, in which the
$\eta^\prime$ is produced through its coupling to gluons. 
We have proposed a  parametrization of
those effects in the $D_s \to \eta^\prime$ form factor. As for 
$D_s \to \eta$, we used a theoretical calculation of the
form factor $f_+^\eta(q^2)$ which corresponds to a branching fraction
for the decay $D_s \to \eta \ell \nu$ in agreement with  data. 
A fit to all the available experimental results,
adopting a generalized factorization scheme for  nonleptonic decays, 
is possible; it  constrains the parameters 
in restricted regions that can be discriminated by making dedicated
observations, for example looking at the semileptonic spectrum of 
the $D_s \to \eta^\prime$ transitions.
An improvement in the precision of the experimental data on $D_s$ decays
could support this scheme and be helpful in understanding 
the dynamics of the $\eta$ and $\eta^\prime$ production in heavy 
meson decays.

\end{document}